\documentclass[a4paper,twoside]{jpconf_mod}
\usepackage{graphicx,amssymb,cite} 

\begin{document}
\title{Constraining super-critical string/brane cosmologies with astrophysical data}

\author{Vasiliki A Mitsou}

\address{Instituto de F\'{i}sica Corpuscular (IFIC), CSIC -- Universitat de Val\`encia, \\ Apartado de Correos 22085, E-46071 Valencia, Spain}
\ead{vasiliki.mitsou@ific.uv.es}

\begin{abstract}
We discuss fits of unconventional dark energy models to the available data from high-redshift supernovae, distant galaxies and baryon oscillations. The models are based either on brane cosmologies or on Liouville strings in which a relaxation dark energy is provided by a rolling dilaton field (Q-cosmology). Such cosmologies feature the possibility of effective four-dimensional negative-energy dust and/or exotic scaling of dark matter. We find evidence for a negative-energy dust at the current era, as well as for exotic-scaling ($a^{-\delta}$) contributions to the energy density, with $\delta\simeq4$, which could be due to dark matter coupling with the dilaton in Q-cosmology models. We conclude that Q-cosmology fits the data equally well with the $\Lambda$CDM model for a range of parameters that are in general expected from theoretical considerations.
\end{abstract}

\section{Introduction}

There is a plethora of astrophysical evidence today, from supernovae
measurements~\cite{HST,davis}, the cosmic microwave
background~\cite{wmap}, baryon oscillations~\cite{baryon} and other cosmological data, indicating that the expansion of the Universe is currently accelerating. The energy budget of the Universe seems to be dominated at the present epoch by a mysterious dark energy component. Many theoretical models provide possible explanations for the latter, ranging from a cosmological constant~\cite{concordance} to super-horizon perturbations~\cite{riotto} and time-varying quintessence scenarios~\cite{steinhardt}, in which the dark energy is due to a smoothly varying scalar field dominating cosmology in the present era. In the context of string theory, such a time-dependent `quintessence' field is provided by the scalar dilaton field of the gravitational string multiplet~\cite{aben,gasperini,emnw}.

\section{Dissipative Q-cosmology basics}

Most of the astrophysical analyses so far are based on effective
four-dimensional Robertson-Walker Universes, satisfying on-shell dynamical
equations of motion of the Einstein-Friedman form. Even in modern approaches to brane cosmology, described by equations deviating during early eras of the Universe from the standard Friedman equation, the underlying dynamics is
assumed to be of classical equilibrium (on-shell) nature.

However, cosmology may not be an entirely classical equilibrium situation.  The initial Big Bang or other catastrophic cosmic event, such as a collision of two brane worlds in the modern approach to strings, which led to the initial rapid expansion of the Universe, may have caused a significant departure from classical equilibrium dynamics in the early Universe, whose signatures may still be present at later epochs including the present era. Q-cosmology is a specific dark energy model, being associated with a rolling dilaton field that is a remnant of this non-equilibrium phase, that was formulated~\cite{emnw,diamandis} in the framework of non-critical string theory~\cite{aben,emn}. 




\section{Data analysis}

We highlight the results~\cite{EMMN,MM} of the confrontation of cosmological models with data on high-redshift supernovae~\cite{davis}, differential galaxy ages~\cite{Hz} and baryon acoustic oscillation~\cite{baryon}. The predictions for the Hubble rate $H(z)$ of the following three cosmological models are investigated: (a) $\Lambda$CDM, a CDM model with a cosmological constant~\cite{concordance}; (b)
the super-horizon model~\cite{riotto}, where the Universe is assumed to
be filled with non-relativistic matter only and there is no dark energy of
any sort; and (c) Q-cosmology~\cite{emnw,diamandis}. 
 

We assume a parametrisation for $H(z)$ in the Q-cosmology framework at late eras, such as the ones pertinent to the data ($0<z<2$), where some analytic approximations are allowed~\cite{EMMN}:
\begin{equation}\label{eq:formulaforfit}
\frac{H(z)}{H_0} = \sqrt{{\Omega }_3 (1 + z)^3 + {\Omega }_{\delta} (1 +
z)^\delta + {\Omega}_2(1 + z)^2}~,~~{\Omega }_3 + {\Omega}_{\delta} +
{\Omega}_2 = 1,
\end{equation}
with the densities $\Omega_{2,3,\delta}$ corresponding to present-day values
($z = 0$). The $a^{-2}$-scaling contribution is a feature of the dilaton relaxation and should not be confused with the spatial curvature contribution. A complete analysis of the non-critical and dilaton
effects, which turn out to be important in the present era after the inclusion
of matter, requires a numerical treatment~\cite{diamandis}. In general, the
three parameters to be determined by the fit are $\Omega_3$, $\Omega_\delta$
and $\delta$. In earlier studies~\cite{EMMN}, a fixed value of $\delta=4$ was assumed for simplicity, whilst a more complete analysis was performed in ref.~\cite{MM}.

\begin{figure}[ht]
\begin{minipage}[b]{0.48\textwidth}
\includegraphics[width=\textwidth]{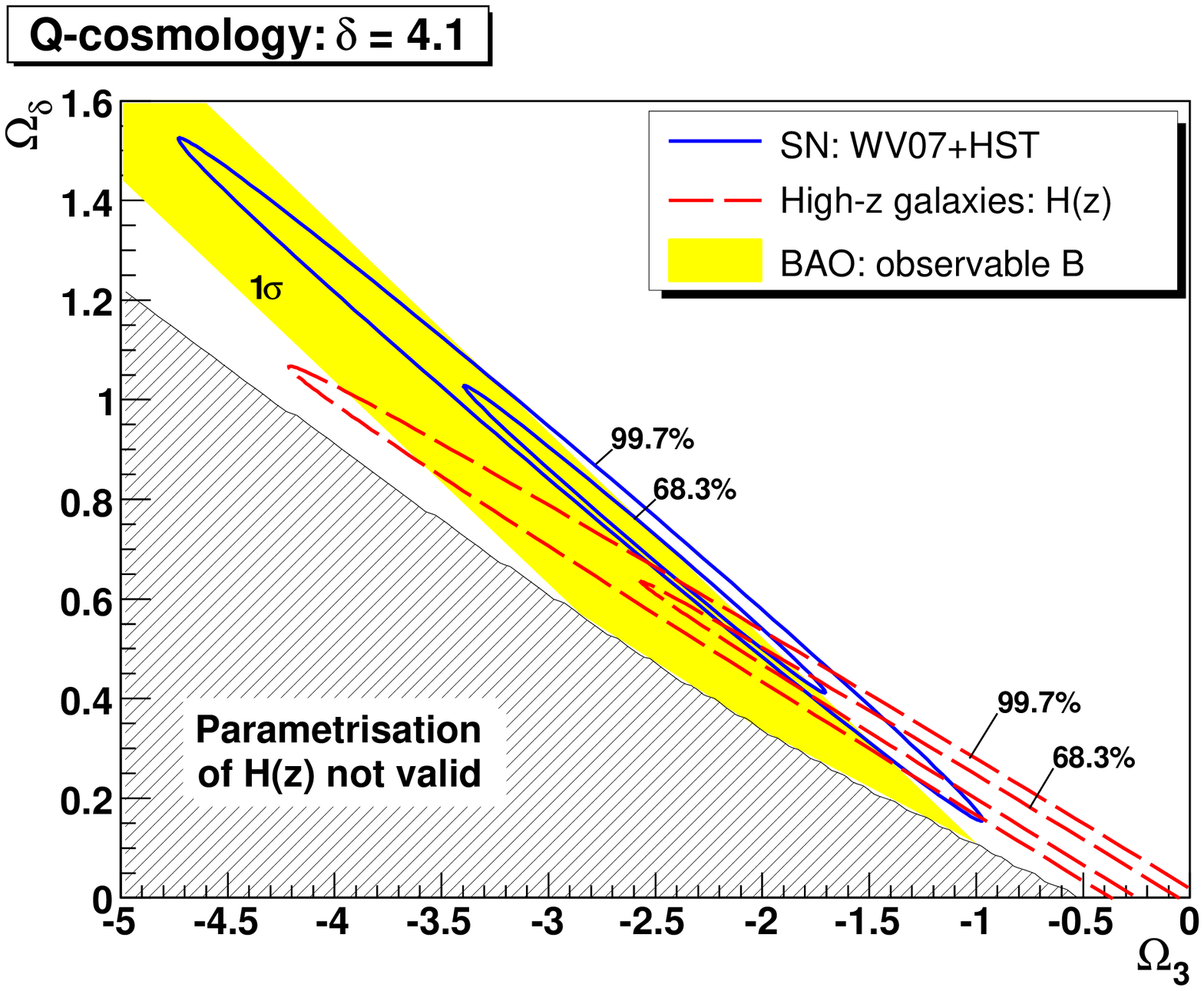}
\caption{\label{fig1}Constraints on Q-cosmology for $\delta = 4.1$: (a) 68.3\% and 99.7\% C.L.\ contours for supernova data (blue solid line); (b) 68.3\% and 99.7\% C.L.\ contours for $H(z)$ (red dashed line); and (c) $1\sigma$ region from the baryon acoustic oscillations (yellow area).}
\end{minipage}\hspace{0.05\textwidth}%
\begin{minipage}[b]{0.47\textwidth}
\includegraphics[width=\textwidth]{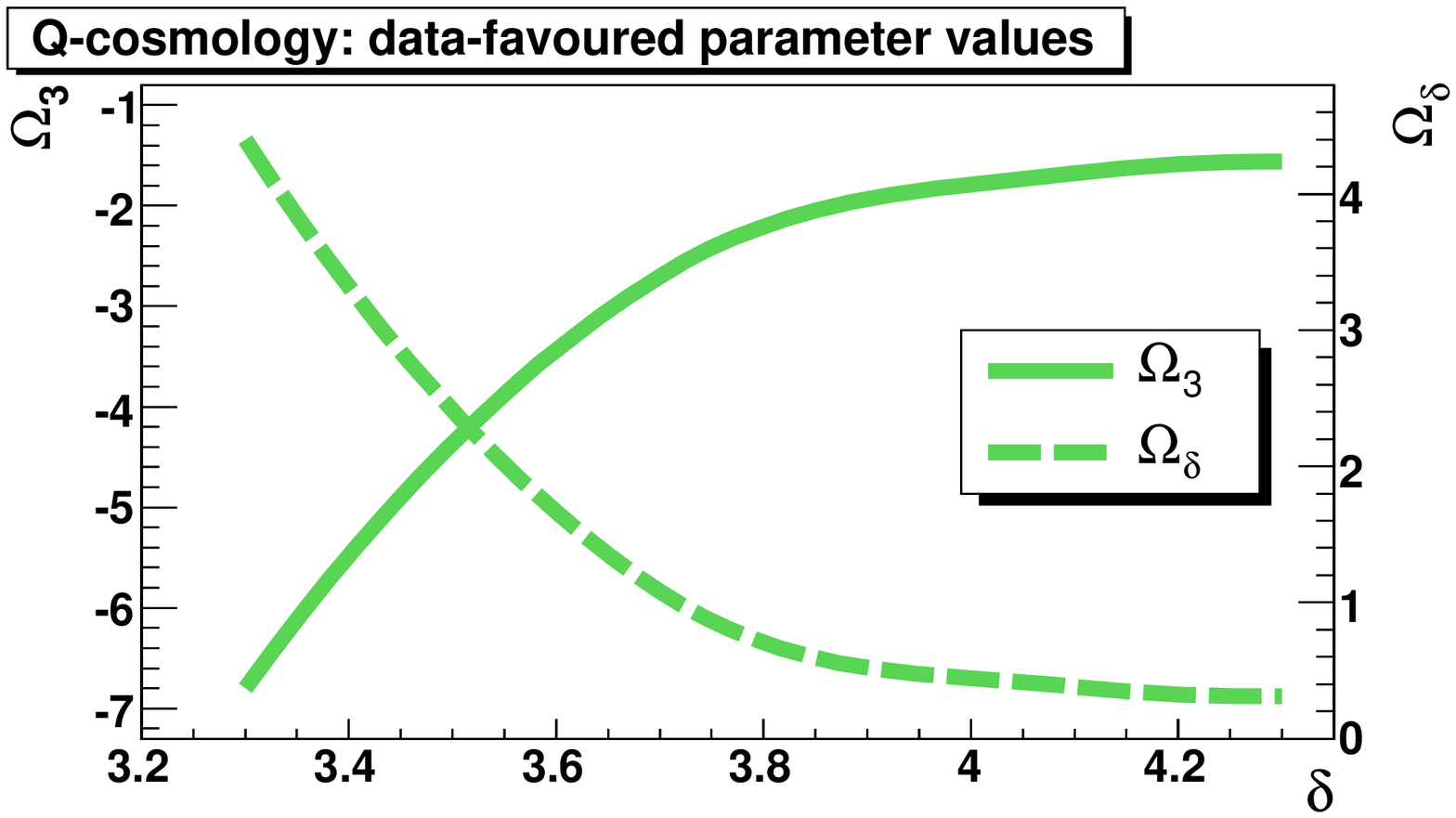}
\caption{\label{fig2} Astrophysical data-favoured values of Q-cosmology parameters as defined from the overlap of the $1\sigma$ regions from supernovae, differential galaxy ages and baryon oscillations data. }
\end{minipage} 
\end{figure}

The $\Lambda$CDM model, as expected, fits the data very well and is mainly constrained by the supernova and BAO data. The super-horizon model, on the other hand, is excluded by the BAO-measured matter density and yields contradictory (at $2\sigma$ level) parameter determinations by SN and galactic data~\cite{MM}.

The spatially-flat Q-cosmology predictions, as parametrised
in~(\ref{eq:formulaforfit}) for $z\lesssim2$, fits the data very well providing an alternative scenario to account for the dark energy component of the Universe. The best-fit region for $\delta=4.1$ is shown in fig.~\ref{fig1}. The three model parameters are not univocally determined; their allowed range and the correlation between them is rather well defined. The parameter $\delta$, associated to the exotic scaling of dark matter in the context of Q-cosmology, is restricted in the range $3.3\lesssim\delta\lesssim4.3$, whilst the present values of (exotic) matter densities, $\Omega_3$ and $\Omega_\delta$, vary with $\delta$ as shown in fig.~\ref{fig2}.

The fact that the value of $\delta=4$ is included in the allowed range, also
points towards dark radiation terms in brane models~\cite{brane}. The data also point towards negative-energy dust contributions, which could be due to either dark-energy dilaton terms in Q-cosmologies~\cite{EMMN,emnw}, or Kaluza-Klein compactification (massive) graviton modes in brane-inspired
modes~\cite{kaluza}.

\section{Other probes of Supercritical String Cosmology (SSC)}

In SSC~\cite{ssc}, D-particle defects in space-time lead to Quantum Gravity `foam«. These point-like stringy defects and the interactions of strings with them lead to non-critical strings, giving rise thus to contributions to Q-cosmology-like situations. The modified dispersion relations in turn produce delays in the propagation of photons. Possible probes of these effects are high energy $\gamma$-rays from distant sources, such as gamma-ray bursts~\cite{farakos}, active galactic nuclei and pulsars. Relevant observations and analyses have been performed by MAGIC~\cite{magic}, HESS~\cite{hess} and Fermi~\cite{fermi1}, with the latter setting the stringent limit from GRB090510~\cite{fermi2}. The role of source effects has to be unfolded in the interpretation of possible time lags observed. 

The presence of the time-dependent dilaton also affects the dark matter relic density calculation, since it modifies the Boltzmann equations. 
A dilution in the neutralino density of ${\cal O}(10)$ is predicted~\cite{lahanas}, widening the allowed parameter space of Supersymmetry at collider searches. The LHC signatures themselves are also affected, favouring final states such as Higgs+jets+MET, Z+jets+MET and $2\tau$+jets+MET, where MET is the missing transverse energy~\cite{dutta}.

\section{Conclusions and outlook}
%

We confirmed~\cite{MM} the results of earlier studies~\cite{EMMN} on high-redshift supernovae data, establishing stringent constraints on cosmological models by analysing, in addition, data from distant galaxies and baryon acoustic oscillations. We demonstrated that cosmological models with no dark energy, such as Q-cosmology, may be viable alternatives to the Standard $\Lambda$CDM model. The data point towards negative-energy dust contributions, which could be due to either dark-energy dilaton terms in Q-cosmologies~\cite{EMMN,emnw}, or Kaluza-Klein compactification (massive) graviton modes in brane-inspired modes~\cite{kaluza}.
%


%

Further detailed studies, such as the theoretical determination of the position of the acoustic peaks in the CMB spectrum, using the underlying formalism of the above models, or the measurement of the pertinent (complicated, in general, $z$-dependent) equation of state are certainly required in order to provide further evidence that might discriminate among the various models. In addition to the CMB spectrum, further constraints on Q-cosmology may be established by data on look-back times of galaxy clusters and by GRBs. 


\ack

This work was conducted in collaboration with Nick Mavromatos, whom I thank for insightful discussions on the theoretical model. The author acknowledges support from the Ram\'on y Cajal contract RYC-2007-00631 of the Spanish Ministry of Science and Innovation (MICINN) and the CSIC. This work was supported in part by MICINN under the projects FPA2006-13238-C02-01 and FPA2006-03081.

\section*{References}

\end{document}